\documentclass[useAMS,usenatbib]{mn2e}

\usepackage{graphics}
\usepackage{epsfig}

\usepackage{longtable}
\usepackage{times}
\title[New Star Clusters]{
Filling the Gap: a New Class of Old Star Cluster?\\
}
\author[Duncan A. Forbes]
{Duncan A. Forbes,$^{1}$\thanks{E-mail: dforbes@swin.edu.au}, 
Vincenzo Pota$^{1}$, Christopher Usher$^{1}$
Jay Strader$^{2}$,  
\newauthor Aaron J.\ Romanowsky,$^{3,4}$, Jean P.\ Brodie$^{4}$, 
Jacob A. Arnold$^{4}$, Lee R. Spitler$^{5,6}$
\\
$^{1}$Centre for Astrophysics and Supercomputing, Swinburne University,
Hawthorn, VIC 3122, Australia\\
$^{2}$Department of Physics and Astronomy, Michigan State University, 
East Lansing, MI 48824, USA\\
$^{3}$Department of Physics and Astronomy, San Jos\'e State University,
One Washington Square, San Jose, CA 95192, USA\\
$^{4}$University of California Observatories, 1156 High St.,
Santa Cruz, CA 95064, USA\\
$^{5}$Department of Physics and Astronomy, Faculty of Sciences, 
Macquarie University, Sydney, NSW 2109, Australia\\
$^{6}$Australian Astronomical Observatory, PO Box 915, North Ryde, NSW 1670, 
Australia\\
}
\begin{document}

%\date{Accepted 1988 December 15. Received 1988 December 14; in original form 1988 October 11}

\pagerange{\pageref{firstpage}--\pageref{lastpage}} \pubyear{2002}

\maketitle

\label{firstpage}

\begin{abstract}

It is not understood whether long-lived star clusters possess a
continuous range of sizes and masses (and hence densities), or
if rather, they should be considered as distinct types with 
different origins.
Utilizing the Hubble Space Telescope (HST) to measure
sizes, and long exposures on the Keck 10m telescope to obtain
distances, we have discovered the first confirmed star clusters 
that lie within a previously claimed
size-luminosity gap dubbed the `avoidance zone' by Hwang et al
(2011). 
The existence of
these star clusters extends the range of
sizes, masses and densities for star clusters, and 
argues against current formation 
models that predict well-defined 
size-mass relationships (such as stripped nuclei, giant
globular clusters or merged star clusters). 
The red colours of these gap objects
suggests that they are not a new class of object but are 
related to Faint Fuzzies observed in nearby lenticular galaxies.
We also report a number of low luminosity UCDs with sizes of up
to 50 pc. 
Future, statistically complete, studies will
be encouraged now that it is known that star clusters possess a continuous
range of structural properties. 
 
\end{abstract}


\section{Introductory Remarks}

%\subsection{Families of old star clusters}
Old, compact star clusters have
traditionally been classified into several types. These include
globular clusters (GCs) first discovered in 1665 by Abraham
Ihle (as noted by Schultz 1866). 
They are compact
(having projected half-light sizes R$_h$ of $\sim$3 pc) and span a wide
range of mass. All large galaxies, including our own Milky Way,
host a system of GCs. 

In the last decade, several new types of star cluster containing
an old stellar population have been
identified.
Deep imaging of the nearby lenticular galaxy NGC
1023 by the Hubble Space Telescope and spectroscopic follow-up using the 10m 
Keck I telescope revealed a population of 
low luminosity GC-like 
objects with large sizes ($\sim$
10 pc) dubbed Faint Fuzzies (FFs) by \cite{LB00}. 
Objects with similar sizes and luminosities were  
discovered around M31 by \citet{H05} and named 
Extended Clusters (ECs). 
%The Milky Way has a dozen GCs with similar properties. In the same time frame,
Similar extended objects have been identified in galaxies ranging from
dwarfs to giant ellipticals (e.g. \cite{P06}, \cite{G09}), and
may be related to the Palomar-type GCs found in the outer halo of
the Milky Way. 

Searches beyond the Local Group have revealed an additional
population of star clusters called 
Ultra Compact Dwarfs (UCDs; Drinkwater et al. 2000). These spherical
collections of stars were first thought to be very compact dwarf
galaxies but they also resemble extended
(R$_h$ $>$ 10 pc) GCs, some two magnitudes  brighter than EC/FFs.  
The origin of these various star clusters (GCs, EC/FFs and UCDs)
and their relationship to each other is the subject of debate 
(e.g. \cite{FK}; \cite{WS}). 
%There are no confirmed objects with masses
%intermediate between those of EC/FFs and UCDs.

The size and luminosity distribution of star clusters was
summarised recently by \cite{B11} (to download database see:\\
http://sages.ucolick.org/downloads/sizetable.txt).
They included all types of
known star cluster with old ($\ge$ 5 Gyr) stellar
ages. They also restricted their sample to objects
with {\it confirmed} distances. This is important if one is
exploring size and luminosity trends, but this has not always been the
case in the literature. 
%Such star clusters are dark
%matter free so the luminosity of a cluster is directly related to its
%mass. 
From size and luminosity, 
the projected surface and volume densities can also be
derived.

%In particular, they included the recently confirmed low
%luminosity UCDs found around the giant elliptical galaxy M87 by
%Strader et al. (2011). 
In Figure 1 we show the fundamental parameters of size and
luminosity from this state-of-the-art compilation for long-lived
star clusters. The figure shows a U-shaped distribution. The high
luminosity, extended star clusters are generally referred to as
UCDs, the base of the U-shape is occupied
by compact GCs and the low luminosity,
extended size regime is associated with ECs and FFs. 
Two extreme Milky Way GCs are
highlighted in the figure; NGC 2419 (the largest Galactic 
GC, which lies in the region
near low luminosity UCDs) and $\omega$ Cen (the most luminous
Galactic GC).
The figure shows that star clusters with V band magnitudes M$_V$
brighter than --10 and projected half-light radii R$_h$ greater
than 5 pc are very rare, if not completely absent, in the Local
Group of galaxies which is dominated by the Milky Way and
Andromeda. Only a few objects beyond the Local Group are known
with M$_V$ fainter than --8.5. This corresponds to an apparent magnitude
limit of V $<$ 22.5 at the distance of the Virgo cluster (a
typical limiting magnitude for spectroscopic studies on 8m class
telescopes). The exception is the deep HST and Keck telescope
observations of FFs in NGC 1023 by 
\cite{LB00}.  The figure also highlights the lack of very
compact, very luminous objects, i.e. those with ultra high
densities. It has been argued by \cite{H10} that feedback from
massive stars sets an upper density limit, beyond which star
clusters do not form.

\begin{figure}
\includegraphics[width=0.48\textwidth,angle=0]{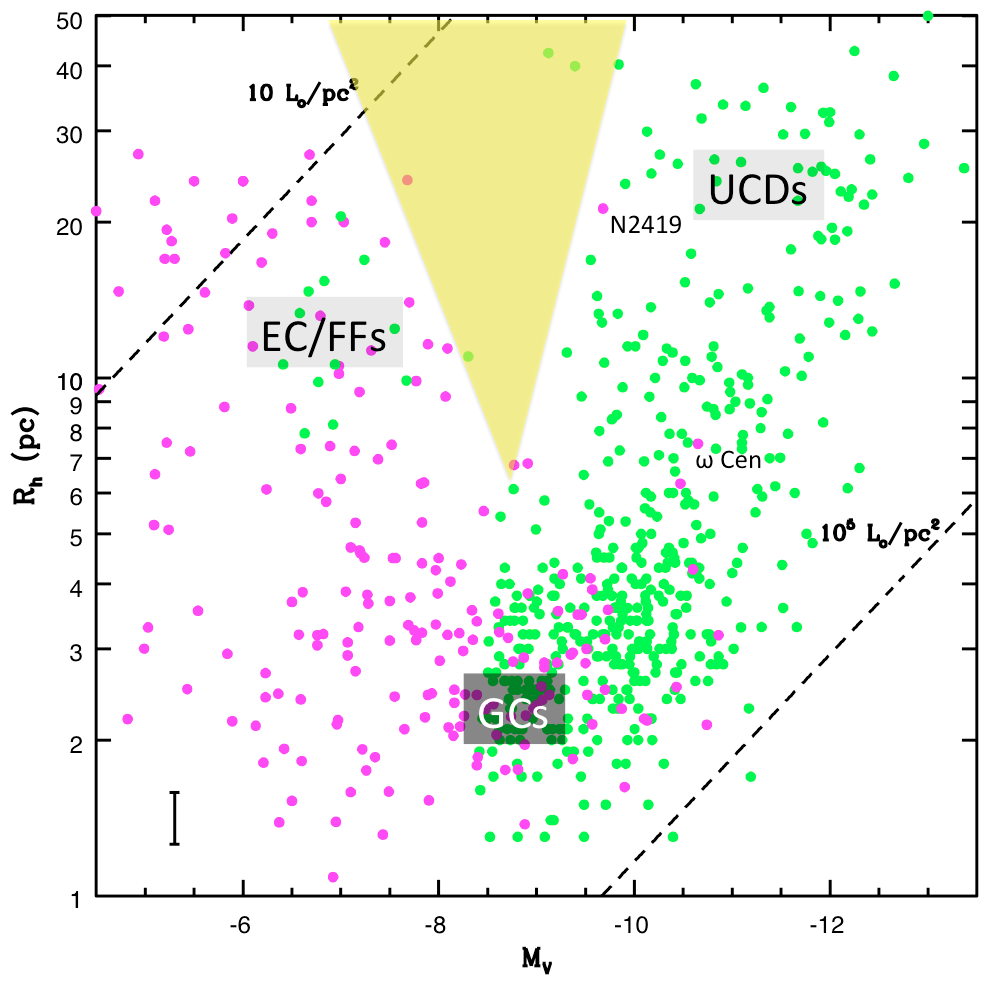}
\caption{Size-luminosity diagram for old star clusters. The
   half-light radius and V-band absolute magnitude for  
   star clusters with known distances and old stellar ages 
from the compilation of Brodie et al. (2011) are shown. The purple symbols denote star
   clusters within the Local Group, while green symbols denote
   star clusters around galaxies beyond the Local
   Group. The general location of Ultra Compact Dwarfs (UCDs),
   Globular Clusters (GCs) and Extended Clusters/Faint Fuzzies
   (ECs/FFs) are labelled, as are the largest Milky Way GC NGC
   2419 and the most luminous one $\omega$ Cen. 
   A typical uncertainty in star cluster size is shown lower left. 
   The diagonal dashed lines denote lines of constant surface
   density, i.e. 10$^{5}$ and 10 solar luminosities per parsec
   squared. Star clusters would be easily
   detectable in the lower right side of this diagram if they
   existed. The upper left portion in this diagram is
   associated with lower densities and lower surface
   brightnesses; hence objects become increasingly difficult to confirm
   observationally. The distribution of known old star clusters shows a
   U-shape 
   with a clear lack of confirmed objects (highlighted by the
   yellow shaded 
region),  
which has been called the star cluster `avoidance zone' by Hwang
et al. (2011). 
}
\end{figure}

However, perhaps the most interesting feature of Figure 1 is the
deficiency of objects around M$_V \approx -9$ and R$_h \ge 7$ pc, i.e. 
sizes and luminosities intermediate between EC/FFs and UCDs.  This
gap in the size-luminosity distribution has been called the star
cluster `avoidance zone' by \cite{H11}. Such a gap
could be due to physical processes or to an observational selection
effect. A real gap would imply that EC/FFs are physically 
distinct from low luminosity UCDs and hence are formed by
different mechanisms that have inherent upper and lower mass limits
respectively. Continuity across the gap might suggest that one
family of star cluster has a wider range of properties than
previously known or that a new type of star cluster exists. 

Here we briefly present the 
recession velocities, and hence physical sizes and luminosities
for extended (R$_h$ $>$ 5pc) star clusters around three
elliptical galaxies. In particular, we investigate whether these
star clusters occupy the `avoidance
zone' seen in Figure 1 or not.

%Obs and data
\section{The Data}

To identify potential star clusters in the
`avoidance zone' the candidates need to be resolved in order to measure their
sizes. This is best achieved with the superior spatial resolution
of
the Hubble Space Telescope (HST).  
A small number of nearby elliptical galaxies have been imaged by
HST in two filters (required for colour selection) and over half
a 
dozen pointings
(needed to identify a large number of candidate star clusters associated
with each galaxy). In particular, half-light sizes have been
measured from g and z band HST/ACS images for candidate star clusters in 
NGC 4278 by Usher et al. (2013, in prep.) and NGC 4649 by 
Strader et al. (2012). In both the Usher et al. and Strader et
al. works, objects were selected on the
basis of having colours that matched those expected of candidate star
clusters. Sizes were then determined using the ISHAPE software
and visual inspection to remove obvious background galaxies. 
For NGC 4697 a similar procedure was used.
The galaxies are located at distances of 15.6 Mpc (NGC 4278), 
17.3 Mpc (NGC 4649) and 11.4 Mpc (NGC 4697). At these distances
HST can resolve sizes as small as 1--2 pc.

After selecting resolved star cluster candidates (with GC-like
colours) around these
three galaxies, we designed 
several multi-object slit masks for the DEIMOS  instrument on the
10m Keck II telescope. Typical exposures of 2 hrs, in 0.8--1.2 arcsec
seeing conditions during the nights of 2013 January 11-12, were
obtained.  
The resulting spectra were reduced using standard procedures
and radial velocities measured, e.g. following the method of 
\cite{P13}. 
For each galaxy we confirmed several tens of GCs, 
with sizes of $\sim$3 pc, to have velocities consistent
with that of their host galaxy. A small number of background
galaxies, with significantly higher velocities, 
were confirmed in each mask. Their magnitudes, colors and angular
sizes of the background galaxies are provided in the Appendix. 

Here we focus on the confirmed objects with sizes greater than 5
pc.
%In Table 1 we present details of the extended (R$_h$ $>$ 5 pc) objects 
%with radial velocities that
%confirm their association with their host galaxy. The host
%galaxies and distances are NGC 4278
Table 1 lists their 
magnitudes, colours, average half-light
radii from the g and z bands and 
apparent V band magnitudes from the transformation:
$0.753 \times (g-z) - 0.108 + z$ (based on a large sample of GCs 
from Usher et al. 2013, in prep.). IDs for the objects come from
Usher et al. (2013, in prep.), Strader et al. (2012) and this
work for NGC 4278, 4649 and 4697 respectively.  
%, where $(m-M)$ is the distance modulus.  

\begin{table}
\caption{Confirmed star clusters with half-light radii
greater than 5 pc}
\begin{tabular}{lccccccc}
\hline\hline
ID & z & err & (g--z) & err & R$_h$ & err & M$_V$ \\
   & (mag) & (mag) & (mag) & (mag) & (pc) & (pc) & (mag) \\
\hline
N4278 & & & & & & & \\
\hline 
acs0320 & 20.78 & 0.030 & 0.845 & 0.041 & 23.23 & 0.51 & -9.65 \\
acs0259 & 20.00 & 0.026 & 0.899 & 0.039 & 21.12 & 0.64 & -10.37 \\
acs0498 & 21.11 & 0.012 & 0.882 & 0.019 & 8.47 & 0.10 & -9.29 \\
acs1362 & 20.30 & 0.027 & 0.810 & 0.030 & 8.35 & 0.67 & -10.10 \\
acs2305 & 21.95 & 0.026 & 1.260 & 0.039 & 6.34 & 0.55 & -8.17\\
acs1369 & 18.43 & 0.014 & 1.440 & 0.021 & 6.07 & 0.34 & -11.55 \\
acs0965 & 21.13 & 0.016 & 0.970 & 0.023 & 5.83 & 0.32 & -9.20 \\
acs1365 & 21.15 & 0.013 & 0.986 & 0.019 & 5.55 & 0.22 & -9.18 \\
acs1606 & 20.59 & 0.014 & 0.906 & 0.020 & 5.53 & 0.30 & -9.80 \\
acs1634 & 22.07 & 0.025 & 0.973 & 0.038 & 5.11 & 0.55 & -8.26 \\
acs1381 & 20.66 & 0.010 & 0.823 & 0.014 & 5.09 & 0.15 & -9.79 \\
\hline
N4649  & & & & & & & \\
\hline
 D68 & 19.66 & 0.062 & 0.991 & 0.087 & 47.42 & 3.37 & -10.78 \\
 A155 & 21.59 & 0.039 & 1.026 & 0.047 & 39.52 & 4.22 & -8.83 \\
 %A155 & 21.59 & 0.039 & 1.026 & 0.047 & 39.52 & 4.22 & -8.83 &  \\
 A32 & 19.91 & 0.018 & 0.923 & 0.023 & 36.48 & 2.64 & -10.57  \\
 E91 & 21.85 & 0.020 & 1.690 & 0.031 & 16.09 & 1.65 & -8.04  \\
 C84 & 21.46 & 0.021 & 1.591 & 0.031 & 14.99 & 1.37 & -8.53  \\
 %C84 & 21.46 & 0.021 & 1.591 & 0.031 & 14.99 & 1.37 & -8.53 &  \\
 C42 & 20.80 & 0.016 & 0.919 & 0.021 & 14.91 & 1.11 & -9.69  \\
 %C42 & 20.80 & 0.016 & 0.919 & 0.021 & 14.91 & 1.11 & -9.69 &  \\
 A122 & 21.73 & 0.027 & 1.600 & 0.040 & 14.69 & 1.51 & -8.25  \\
 %A122 & 21.73 & 0.027 & 1.600 & 0.040 & 14.69 & 1.51 & -8.25 &  \\
 A51 & 21.10 & 0.016 & 1.659 & 0.025 & 13.79 & 1.14 & -8.83  \\
 %A51 & 21.10 & 0.016 & 1.659 & 0.025 & 13.79 & 1.14 & -8.83 &  \\
 C28 & 20.58 & 0.015 & 0.927 & 0.019 & 13.58 & 0.99 & -9.90  \\
 %C28 & 20.58 & 0.015 & 0.927 & 0.019 & 13.58 & 0.99 & -9.90 &  \\
 B139 & 21.71 & 0.021 & 0.952 & 0.028 & 11.17 & 0.96 & -8.76  \\
 J67 & 19.81 & 0.019 & 0.877 & 0.031 & 9.83 & 0.17 & -10.72  \\
 %D3 & 16.53 & 0.010 & 1.522 & 0.014 & 9.34 & 0.66 & -13.52  \\
 J623 & 20.93 & 0.049 & 0.934 & 0.065 & 6.88 & 0.26 & -9.55  \\
 B8 & 18.62 & 0.010 & 1.314 & 0.015 & 5.58 & 0.39 & -11.59  \\
 %B8 & 18.62 & 0.010 & 1.314 & 0.015 & 5.58 & 0.39 & -11.59 &  \\
 J76 & 19.19 & 0.022 & 1.230 & 0.042 & 5.55 & 0.15 & -11.08  \\
 A197 & 22.76 & 0.031 & 1.085 & 0.041 & 5.04 & 0.72 & -7.61
 \\
\hline
N4697 & & & & & & & \\
\hline
acs52 & 19.25 & 0.002 & 0.846 & 0.004 & 27.30 & 2.59 & -10.49 \\
acs580 & 22.09 & 0.017 & 1.342 & 0.059 & 26.42 & 8.43 & -7.27 \\
acs112 & 20.05 & 0.004 & 0.882 & 0.006 & 19.15 & 1.92 & -9.67 \\
acs173 & 20.48 & 0.005 & 1.021 & 0.008 & 19.18 & 1.98 & -9.13 \\
acs132 & 20.24 & 0.004 & 0.865 & 0.007 & 15.57 & 2.25 & -9.49 \\
acs486 & 21.79 & 0.012 & 0.960 & 0.017 & 7.85 & 3.03 & -7.86 \\
acs071 & 19.60 & 0.003 & 0.906 & 0.005 & 5.37 & 2.14 & -10.10 \\
acs474 & 21.75 & 0.010 & 0.960 & 0.016 & 4.90 & 1.98 & -7.91 \\
acs782 & 22.59 & 0.021 & 1.515 & 0.039 & 7.08 & 3.14 & -6.65 \\
acs270 & 20.99 & 0.008 & 1.300 & 0.012 & 5.29 & 1.00 & -8.41 \\
acs607 & 22.18 & 0.016 & 1.445 & 0.029 & 6.14 & 2.81 & -7.11 \\
acs805 & 22.67 & 0.019 & 1.329 & 0.035 & 5.15 & 1.04 & -6.71 \\
\hline
\end{tabular}
\end{table}

\section{Filling the gap}

In Figure 2 we again show the data points from \cite{B11} and now
include
all the confirmed star clusters in NGC 4278, 4649 and 4697. 
%for which we were recently able to 
%confirm their distance (see Section 2). 
Our main finding is that old star
clusters do indeed occupy the `avoidance zone' gap. 
The avoidance zone is therefore simply the result of a selection bias
in previous works which were unable to reach low enough surface
brightness levels beyond the Local Group. 
%Our deep exposures on the
%world's largest telescope have allowed us to 
Here we confirm that long-lived
star clusters cover a wide and continuous range of sizes and
luminosities (and hence densities).
%As most stars orginally form
%in star clusters (Lada \& Lada ***) our findings ...

A clue to the
nature of the extended size star clusters comes from their instrinsic 
colours. In Figure 2 objects have been coded by their colour,
i.e. red or blue for a colour separation at (g--z) = 1.1, which
corresponds to a metallicity of [Fe/H] $\sim$ --1.
We find that the high luminosity star clusters tend to
be blue (or metal-poor) and the low luminosity ones red
(metal-rich). Focusing on the gap itself, the clusters are mostly 
red in colour indicating that they are metal-rich. 
This suggests that they are more closely related to the 
lower luminosity FFs found in NGC 1023 by \cite{LB00} and
the Diffuse Star Clusters (DSCs) of \cite{P06} which are
metal-rich and red in colour. 
These objects are typically associated with the disks of lenticular galaxies
that reveal signs of a past interaction. Burkert, Brodie \&
Larsen (2005) suggest that FFs form in metal-rich disks as the result of an
interaction and subsequent starburst. 
Goudfrooij (2012) has argued that the intermediate-aged diffuse 
star clusters in the merger remnant NGC 1316 may evolve to
resemble FFs after the continued disruption by tidal shocks. 
Although all three host galaxies studied here are classified as ellipticals,
we note that NGC 4278 contains a large HI ring (Raimond et
al. 1981) that is perhaps a remnant of a past interaction, NGC
4649 reveals strong rotation in its outer region as might
expected after a major merger (Hwang et al. 2008) and
NGC 4697 is highly flattened (E6) and so may be a mis-classified
S0 (Dejonghe et al. 1996).

\begin{figure}
\includegraphics[width=0.48\textwidth,angle=0]{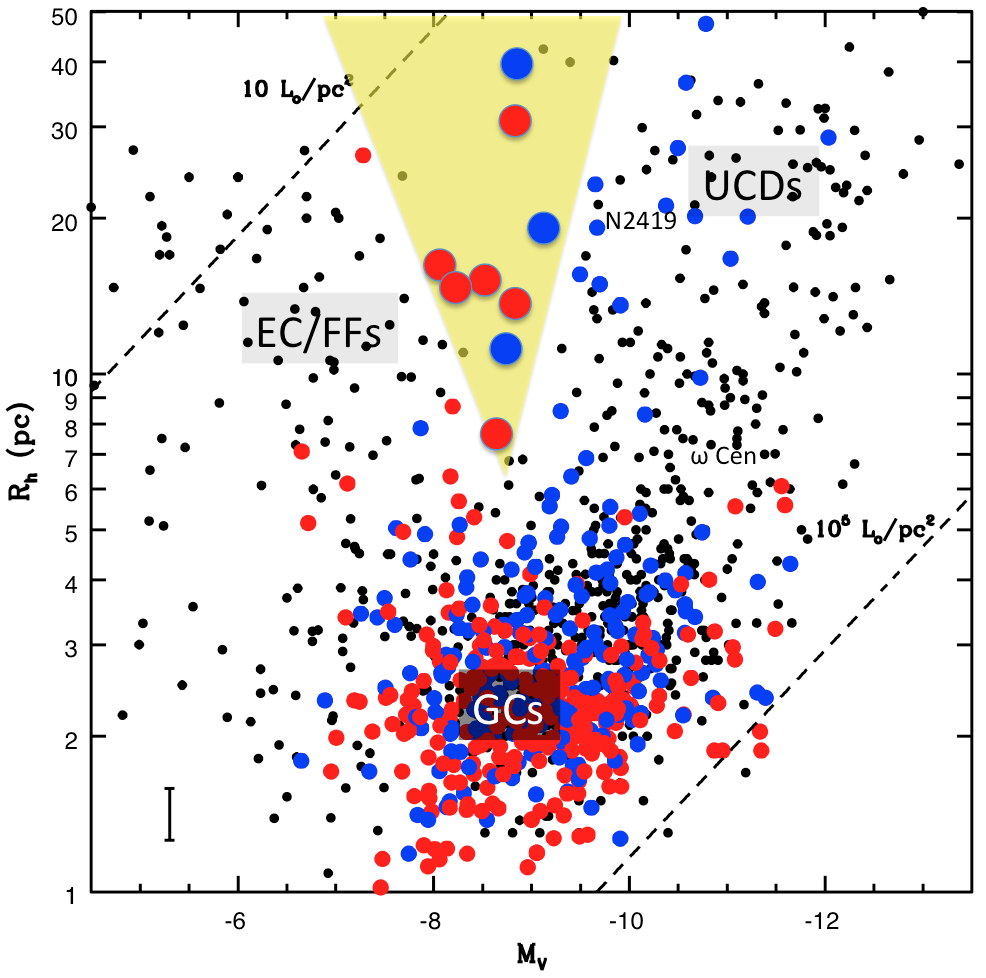}
\caption{Size-luminosity diagram for old star clusters including
   newly confirmed objects around the early-type galaxies NGC
   4278, 4649 and 4697. The data of Brodie et al. (2011) from
   Figure 1 are shown as small black dots. 
Labels are as in
   Figure 1.
New star cluster measurements (with sizes from
   the Hubble Space Telescope and distances from the Keck
   telescope) are shown
 as blue and red symbols (corresponding to a division at 
colour (g--z) = 1.1, equivalent to metallicity [Fe/H] = -1). Several
   new objects, with the largest symbols, occupy the yellow shaded `avoidance
   zone' of Figure 1; thus long-lived star clusters cover a wide
   and continuous range in size and luminosity. The new data also
   include two objects with sizes and luminosities
   similar to the Milky Way GC NGC 2419, several additional objects that
   might be classified as low luminosity UCDs, one very low
   density object that appears to be similar to an extended
   cluster and/or faint fuzzy, an object of similar luminosity
(and hence mass) to $\omega$ Cen but six times larger, as well
   as numerous compact globular clusters. Most of the extended
high luminosity objects are intrinsically blue while the low luminosity ones,
including those in the `avoidance zone' gap, tend to be red.
}
\end{figure}

We have also confirmed the existence of 
several other interesting objects. They
include a
number of blue low luminosity UCDs, similar to those
found originally by \cite{S11} and listed in their table 9. Two of these have 
sizes and luminosities very similar to 
the Milky Way GC NGC 2419, the largest known GC
in the Milky Way. 
%It's size is not due to internal relaxation
%processes as the half mass relaxation time is several times the
%age of the Universe (Ibata et al. 2013). 
Like other massive GCs in the Milky Way, NGC 2419 contains multiple stellar
populations, e.g. \cite{CK12}, which are traditionally associated with
galaxies (\cite{FK}; \cite{WS}). Indeed \cite{CK12} have suggested
that NGC 2419 is not in fact a GC but the remnant nucleus of a
stripped dwarf galaxy.
If it was
once part of a dark matter dominated dwarf galaxy, that dark
matter appears to have been largely stripped away as none is detected
today in its outer regions (\cite{C11}; \cite{I13}). 

%We also have several objects that border the region of ECs/FFs. The FFs
%in NGC 1023 are known to be red and presumably metal-rich. Figure
%2 shows several red objects with similar sizes but extending to
%higher luminosities than the NGC 1023 FFs. 
We also confirm
an object (acs580) around NGC 4697 with a similar luminosity to the red
FFs but 
with a larger
size (26 pc). This object has the lowest surface density
of any confirmed old star cluster beyond the Local Group. 
Finally, we note that one star cluster (D68) 
has a luminosity of M$_V$ =  --10.8, similar to that of $\omega$ Cen
(the most massive GC or
remnant nucleus in
the Milky Way) but with a half light radius some six times larger 
at 47 pc and hence a lower surface density by a factor of $\sim$35.

\begin{figure}
%\begin{center}
\includegraphics[width=0.48\textwidth,angle=0]{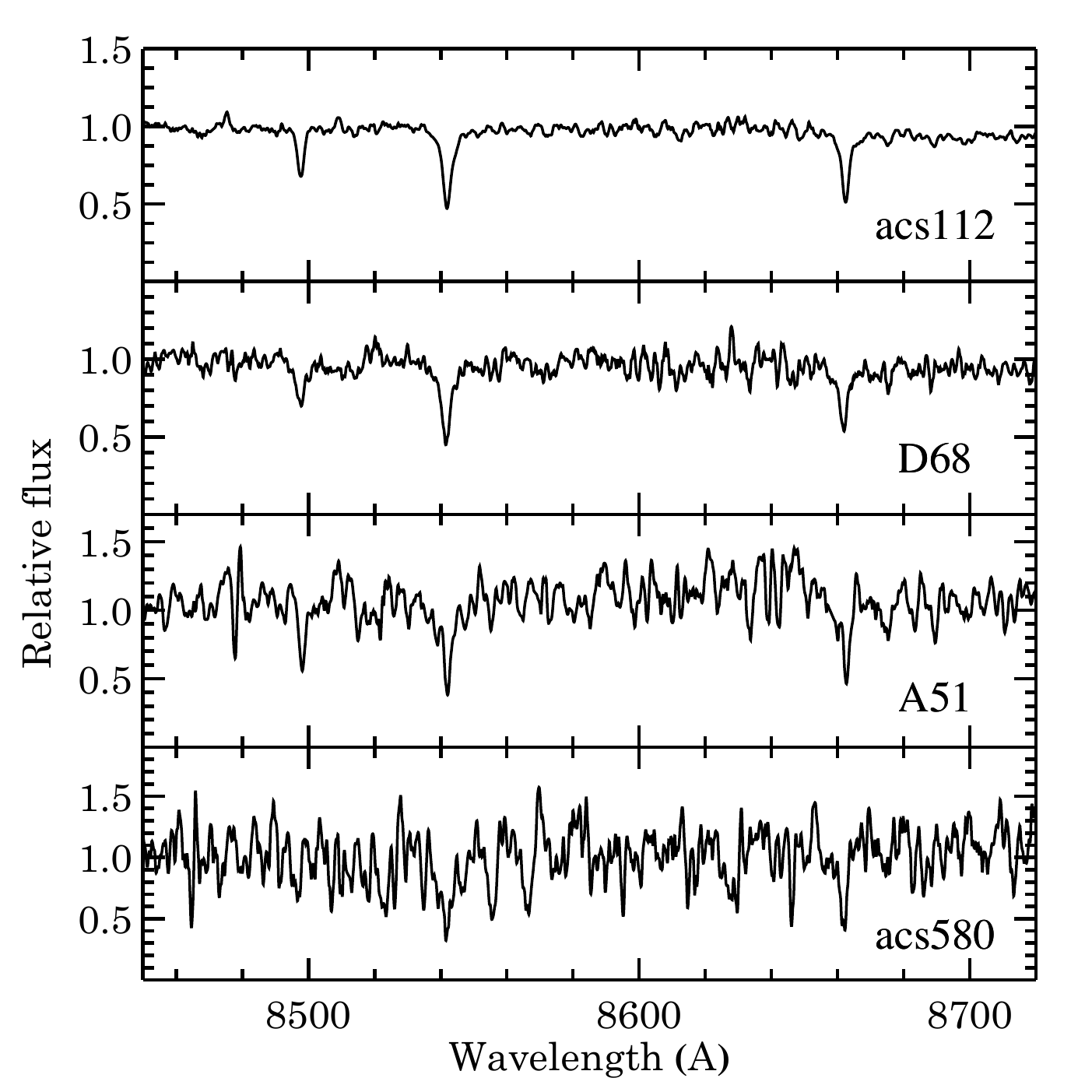}
 \caption{Keck spectra of selected star clusters. The three
Calcium Triplet lines used for redshift determination are visible
near 8498, 8542 and 8662~\AA. The plot shows
from top to bottom in decreasing signal-to-noise: 
objects acs112 associated with NGC 4697 (M$_V$ = --9.67, R$_h$ = 19
pc), D68 in NGC 4649 (M$_V$ = --10.78, R$_h$ = 47 pc),
A51 in NGC 4649 (M$_V$ = --8.83, R$_h$ = 14 pc) and acs580 
in NGC 4697 (M$_V$ = --7.27,
R$_h$ = 26 pc). acs112 has a size and luminosity similar to that of the
Milky Way GC NGC 2419. D68 is
the largest star cluster confirmed in this work.
A51 is a red star cluster located in the
`avoidance zone'. acs580 is a large Faint Fuzzy (FF) analogue in NGC
4697. 
}
%\end{center}
\end{figure}

In Figure 3 we show several examples of our Keck spectra for
selected star clusters. The examples include: 
acs112 which has a size and luminosity similar to that of the
Milky Way GC NGC 2419; D68, 
the largest star cluster confirmed in this work;  
%It has a similar
%luminosity to Omega Cen in the Milky Way but is some six times
%larger in size. 
A51, a red star cluster located in the
`avoidance zone' and acs580 a large faint fuzzy (FF) analogue around NGC
4697.

We remind the reader that it becomes
increasingly difficult with decreasing brightness and increasing size 
to confirm low density star clusters (the
upper left hand side of Figure 2) 
and so a reduction in the number of star
clusters in that region of the figure is probably due to current observational
limitations. Future deep surveys may rectify this, finding that region is well-populated.

%Theory
\section{Concluding Remarks}

A number of theories have been put forward to explain the origin
of the different types of extended star cluster with 
corresponding predictions for their structural properties. 
For example, if UCDs are 
simply giant GCs (\cite{M09}) 
or the remnant nuclei of stripped dwarf
galaxies (\cite{B01}) then a well-defined size-luminosity
trend of near constant density is predicted. In the merging star cluster simulations of
\cite{B04} the resulting UCDs are also 
predicted to have a well-defined
size-luminosity relationship. 
Although a distinct size-luminosity relation may exist for more luminous objects (such as compact ellipticals), for luminosities fainter than M$_V$ = --13.5 we find a continuous range in size and 
luminosity for old star clusters. 
%We note that none of these models predicted star
%clusters with sizes and luminosities within the avoidance zone. 
With the introduction of an external tidal field, and exploring a
larger range of masses, the simulations of \cite{BK11} produced merged
star clusters with a large range of size and luminosity. However,
their work indicated an upper limit to the maximum size that increased
with star cluster mass. This is not generally seen in our data.

Individual star clusters were assumed to follow a distinct
initial 
size-mass relationship in the simulations of \cite{G10} but the
effects of stellar evolution, binaries and two-body relaxation over
time resulted in their old clusters having large (R$_h$ $\sim$ 10 pc)
sizes.  Tidal effects would tend to reduce this size further. While
matching some aspects of our data, this model has difficulty
reproducing the largest (R$_h$ $>$ 10 pc) star clusters.

In summary, we find a continuity of
structural properties across a gap in size and luminosity 
called the `avoidance zone'. The red colours of these gap objects
suggests that they are not a new class of object but are 
related to the Faint Fuzzies observed in nearby lenticular galaxies.
We also report a number of low luminosity UCDs with sizes of up
to 50 pc. 
%The lowest luminosity UCDs, such as NGC
%2419, tend to be blue (metal-poor). 
No single model for the formation of extended star clusters can
currently reproduce the diversity of structural properties now observed for old
star clusters.\\

\noindent
{\bf ACKNOWLEDGEMENTS}\\

\noindent
The data presented herein were obtained at the W.M. Keck
  Observatory, which is operated as a scientific partnership
  among the California Institute of Technology, the University of
  California and the National Aeronautics and Space
  Administration. The Observatory was made possible by the
  generous financial support of the W.M. Keck Foundation. The
  analysis pipeline used to reduce the DEIMOS data was developed
  at UC Berkeley with support from NSF grant AST-0071048. Based on
  observations made with the NASA/ESA Hubble Space Telescope,
  obtained from the data archive at the Space Telescope Science
  Institute. STScI is operated by the Association of Universities
  for Research in Astronomy, Inc. under NASA contract NAS
  5-26555. DF thanks the ARC for support via DP130100388.  JB 
acknowledges support from NSF grant AST-1109878. We thank the
referee for several useful suggestions that have improved the
paper.

%Figure 2 shows that objects with sizes R$_h$
%$>$ 10 pc exist from luminosities of M$_V$ $\sim$ --13.5 (UCDs) down to
%--4.5 (EC/FFs) without such a trend. This doesn't rule out the
%possibility that some UCDs formed in this way. 

%Here we have only sampled a small fraction of the star cluster systems around three galaxies so 
%to probe the true distribution of star cluster sizes and luminosities a 
%complete volume limited sample is required.
%This will provide valuable additional insight as to their origin.

%\bibliography{ref}

\appendix
\section{Background objects}

We caution that some researchers have attempted to explore the
size-luminosity distribution of star clusters without having a
confirmed distance to each object. This is a dangerous practice
and can lead to incorrect conclusions. For example, some have
been tempted to explore mean size trends with luminosity and to
make subsequent comparisons with theoretical predictions. In  
Table A1 we list the objects which
have similar apparent sizes and magnitudes to our confirmed
objects but our spectroscopic redshifts indicate that they are actually 
distant background galaxies. The columns are ID, z
magnitude and error, (g--z) colour and error, half-light radius and
error.

\begin{table}
\caption{Background galaxies}
\begin{tabular}{lcccccc}
\hline\hline
ID & z & err & (g--z) & err & R$_h$ & err \\
   & (mag) & (mag) & (mag) & (mag) & (arcsec) & (arcsec) \\
\hline
N4278  & & & & & & \\
\hline
acs2464 & 22.18 & 0.02 & 0.84 & 0.03 &  0.07 & 0.003\\
acs0284 & 21.86 & 0.05 & 1.49 & 0.08 & 0.14 & 0.017\\
\hline
N4649  & & & & & & \\
\hline
A78	&	21.14	& 0.020	& 1.56	& 0.028	& 0.23	& 0.026\\		
E86	&	21.68	& 0.023	& 0.61	& 0.029	& 0.18	& 0.015\\		
E123	& 	22.14	& 0.029	& 1.85	& 0.048	& 0.17	& 0.025\\		
A221	& 	22.48	& 0.033	& 1.33	& 0.048	& 0.17	& 0.024\\		
\hline
N4697  & & & & & & \\
\hline
acs836	&	22.76	& 0.022	& 0.92 & 0.022 & 0.19 & 0.009 \\
acs868	&	22.84	& 0.023	& 1.49 & 0.030 & 0.15	& 0.009\\
\hline
\end{tabular}
\end{table}

\end{document}